\pdfoutput=1
\documentclass{article}

\usepackage[final]{nips_2016}

\usepackage[utf8]{inputenc} % allow utf-8 input
\DeclareUnicodeCharacter{2010}{-}
\usepackage[T1]{fontenc}    % use 8-bit T1 fonts
\usepackage{hyperref}       % hyperlinks
\usepackage{url}            % simple URL typesetting
\usepackage{booktabs}       % professional-quality tables
\usepackage{amsfonts}       % blackboard math symbols
\usepackage{microtype}      % microtypography
\usepackage{amsmath}
\usepackage{listings}
\usepackage{wrapfig}
\usepackage[pdftex]{graphicx}
\usepackage[font=small,labelfont=bf]{caption}
\PassOptionsToPackage{numbers,compress}{natbib}
\usepackage{color}

\newcommand{\beginsupplement}{%
        \setcounter{table}{0}
        \renewcommand{\thetable}{S\arabic{table}}%
        \setcounter{figure}{0}
        \renewcommand{\thefigure}{S\arabic{figure}}%
     }

\title{Feedback GAN (FBGAN) for DNA: a Novel Feedback-Loop Architecture for Optimizing Protein Functions}

\author{ Anvita Gupta, James Zou\\
  \texttt{avgupta@stanford.edu, jamesz@stanford.edu} \\
  Department of Biomedical Data Science\\
  Department of Computer Science\\
  Stanford University\\
}

\begin{document}
\maketitle

\begin{abstract}
Generative Adversarial Networks (GANs) represent an attractive and novel approach to generate realistic data, such as genes, proteins, or drugs, in synthetic biology. Here, we apply GANs to generate synthetic DNA sequences encoding for proteins of variable length. We propose a novel feedback-loop architecture, called Feedback GAN (FBGAN) to optimize the synthetic gene sequences for desired properties using an external function analyzer. The proposed architecture also has the advantage that the analyzer need not be differentiable. We apply the feedback-loop mechanism to two examples: 1) generating synthetic genes coding for antimicrobial peptides, and 2) optimizing synthetic genes for the secondary structure of their resulting peptides. A suite of metrics demonstrate that the GAN generated proteins have desirable biophysical properties. The FBGAN architecture can also be used to optimize GAN-generated datapoints for useful properties in domains beyond genomics.
\end{abstract}

\section{Introduction}

Synthetic biology refers to the systematic design and engineering of biological systems, and is a growing domain which promises to revolutionize areas such as medicine, environmental treatment, and manufacturing \cite{Benner2005}. However, current technologies for synthetic biology are mostly manual and require a significant amount of domain experience. Artificial Intelligence (AI) can transform the process of designing biological molecules by helping scientists leverage large existing genomic and proteomic datasets; by uncovering patterns in these datasets, AI can help scientists design optimal biological molecules. In addition, generative models, such as Generative Adversarial Networks (GANs) can automate the process of designing DNA sequences, proteins, and additional macromolecules for usage in medicine and manufacturing.

Solutions for using GANs for synthetic biology require a framework not only for the GAN to generate novel sequences, but also to optimize the generated sequences for desired properties such as binding affinity of the sequence for a particular ligand, or secondary structure of the generated macromolecule. These properties are necessary for the synthetic molecules to posses so they can be useful in various real-world use cases.

Here, we present a novel feedback loop mechanism for generating DNA sequences using a GAN and then optimizing these sequences for desired properties using a separate predictor, which we call a function analyzer.

The proposed feedback loop mechanism is applied to train a GAN to generate protein-coding sequences (genes), and then enrich the produced genes for those that produce 1) antimicrobial peptides, and 2) alpha-helical peptides. Antimicrobial peptides (AMPs) are typically lower molecular weight peptides with broad antimicrobial activity against bacteria, viruses, and fungi \cite{AMP}. They are an attractive area to apply GANs to since they are commonly short, less than 50 amino acids, and have promising applications to fighting drug resistant bacteria \cite{AMP_length}.

Similarly, optimizing for resulting secondary structure of the genes is possible since common secondary structures, such as helices and beta sheets, arise even in short peptides. Secondary structure is also important to when designing proteins for particular functions.

Optimizing for these two properties provides a proof of concept that the proposed feedback-loop architecture FBGAN can be used to effectively optimize a diverse set of properties, regardless of whether a differentiable analyzer is provided for that property.

\paragraph{Related works}
Besides GANs, Recurrent Neural Networks (RNNs) have also shown promise in producing sequences for synthetic biology applications. RNNs have shown to be successful in generating SMILES sequences for \textit{de novo drug discovery} \cite{segler} and recent work also showed that the RNN's outputs could be optimized for specific properties through transfer-learning and fine-tuning on sequences with desired properties \cite{Gupta2017}. A similar methodology has been applied to generate antimicrobial peptides \cite{muller_amp}. RNNs have also been combined with reinforcement learning to produce sequences optimized for certain properties in synthetic biology \cite{olivecrona}.

However, GANs have the attractive property over RNNs that they allow for latent space interpolation with the input codes provided to the generator \cite{DBLP:journals/corr/SalimansGZCRC16}. Indeed, GANs are increasingly being used to generate realistic biological data. Recently, GANs have been used to morphologically profile cell images \cite{Goldsborough227645}, to generate time-series ICU data \cite{Esteban2017RealvaluedT}, and to generate single cell RNA-seq data from multiple cell types \cite{Ghahramani262501}. GANs have also been used to generate images of cells imaged by fluorescent microscopy, uniquely using a separable generator where one channel of the image was used as input to generate another channel \cite{osokin_gan}.

In independent and concurrent work, Killoran \textit{et al.} use GANs to generate generic DNA sequences \cite{duvenaud_dna}. This work used a popular variant of the GAN known as the Wasserstein GAN, which optimizes the earth mover distance between the generated and real samples \cite{wgan}. In this approach, the generator was first pretrained to produce DNA sequences, and then the discriminator was replaced with a differentiable analyzer. The analyzer in this approach was a deep neural network that predicted, for instance, whether the input DNA sequence bound to a particular protein. By backpropagating through the analyzer, the authors modified the input noise into the generator into specific codes to yield desirable DNA sequences. This approach does not extend to nondifferentiable analyzers, and does not change the generator itself, but rather its input.

Here, we propose a novel feedback-loop architecture, FBGAN, to enrich a GAN's outputs for user-desired properties; the architecture employs an external predictor for the desired property which, as an added benefit, need not be differentiable. We present a proof-of-concept of the feedback-loop architecture by first generating realistic genes, or protein-coding DNA sequences, up to 50 amino acids in length (156 nucleotides); feedback is then used to enrich the generator for genes coding for AMPs, and genes coding for alpha-helical peptides.

\section{Methods}
\subsection{GAN Model Architecture}

The basic formulation of a GAN as proposed by Goodfellow \textit{et al} consists of two component networks, a Generator $G$ and a Discriminator $D$, where the generator $G$ creates new data points from a vector of input noise $z$, and the discriminator $D$ classifies those data points as real or fake \cite{GoodfellowGan}. The end goal of $G$ is to produce data points so realistic that $D$ is unable to classify them as fake. Each pass through the network includes a backpropagation step, where the parameters of $G$ are improved so the generated data points appear more realistic. $G$ and $D$ are playing a minimax game with the following loss function \cite{GoodfellowGan}:

\begin{equation}
	\underset{G}{\text{min}} \underset{D}{\text{max}} V(D,G) = \mathbf{E}_{x \in P_{data}(x)} [log(D(x)] + \mathbf{E}_{z \in P(z)} [log(1-D(G(z))]
\end{equation}

Concretely, the discriminator seeks to maximize the probability $D(x)$ that $x$ is real when $x$ comes from a distribution of real data, and minimize the probability that the data point is real, $D(G(z))$, when $G(z)$ is the generated data.

The Wasserstein GAN (WGAN) is a variant of the GAN which instead minimizes the Earth Mover (Wasserstein) distance between the distribution of real data and the distribution of generated data \cite{wgan}. A gradient penalty is imposed for gradients above one in order to maintain a Lipshitz constraint \cite{wgan_gp}.

WGANs have been shown empirically to be more stable during training than the vanilla GAN formulation. Moreover, the Wasserstein distance corresponds well to the quality of the generated data points \cite{wgan}.

Our GAN model for producing gene sequences follows the WGAN architecture with gradient penalty which proposed by Gulrajani \textit{et al} \cite{wgan_gp}. The model has five residual layers with two 1-D convolutions of size $5 \times 1$ each. However, we replace the softmax in the final layer with a Gumbel Softmax operation with temperature $t=0.75$. When sampling from the generator, the argmax of the probability distribution is taken to output a single nucleotide at each position. The model was coded in Pytorch and initially trained for $70$ epochs with a batch size $B=64$.

\subsubsection{GAN Dataset}

A diverse training set of genes was assembled in order to train the GAN to produce protein-coding sequences. More than 3655 proteins were collected from the Uniprot database, where each protein was less than 50 amino acids in length \cite{uniprot}. These proteins were selected from the set of all reviewed proteins in Uniprot with length from 5-50 residues, and the protein sequences were then clustered by sequence similarity $\geq 0.5$. One representative sequence was selected from each cluster to form a diverse dataset of short peptides. The dataset was limited to proteins up to 50 amino acids in length since this length allows for observations of protein properties such as secondary structure and binding activity, while limiting the long-term dependencies the GAN would have to learn to generate sequences coding for these proteins.

The Uniprot peptides were then converted into cDNA sequences by translating each amino acid to a codon (where a random codon was selected when multiple codons mapped to one amino acid); the canonical start codon and a random stop codon were also added to each sequence. All sequences were padded to length 156, which was the maximum possible length.

\subsection{Feedback-Loop Training Mechanism}

\begin{figure}[]
 \begin{minipage}{\textwidth}
 a)\\
    \begin{center}
    \includegraphics[width=0.8\textwidth]{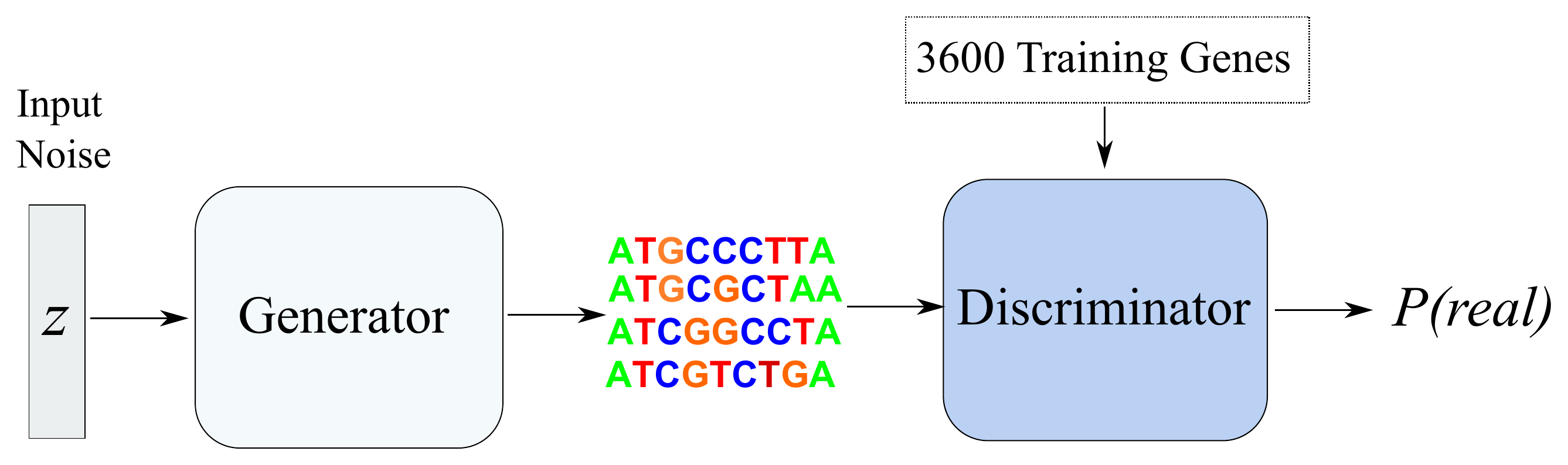}
    \end{center}
  \end{minipage}
   \begin{minipage}{\textwidth}
 b)\\
    \begin{center}
    \includegraphics[width=0.5\textwidth]{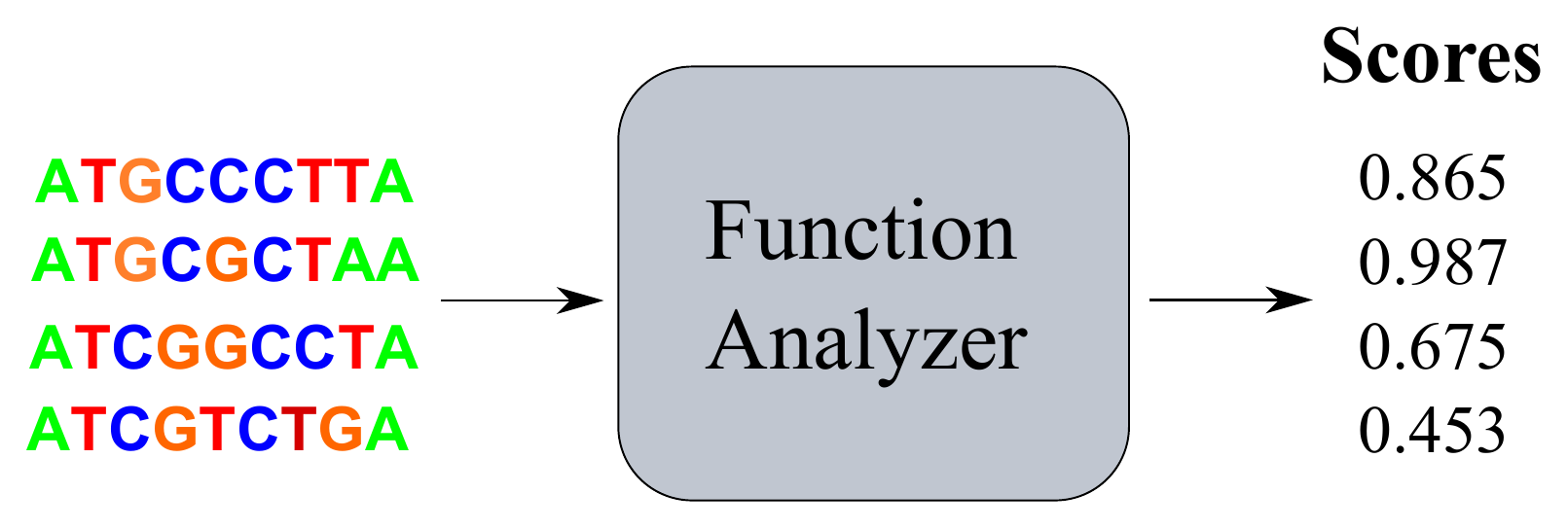}
    \end{center}
  \end{minipage}
  \begin{minipage}{\textwidth}
    c)\\
    \begin{center}
    \includegraphics[width=0.8\textwidth]{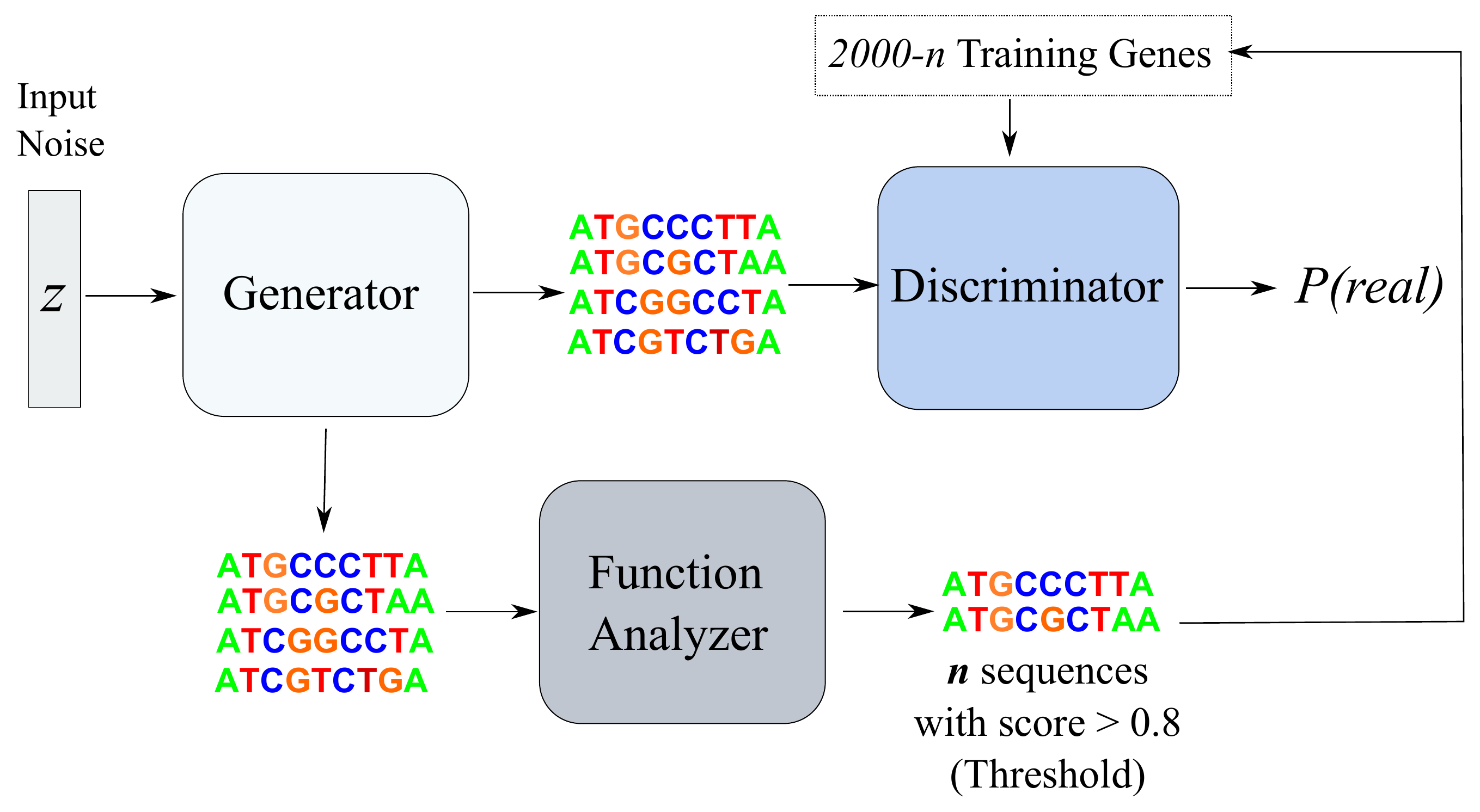}
    \end{center}
  \end{minipage}
  \caption{a) The general training mechanism of the GAN model used to produce gene sequences. The training set used was 3600 genes coding for Uniprot proteins under 50 amino acids in length. b) The general form of the function analyzer, which takes in a sequence and produces a score. The analyzer may be any model which fits this framework, from a deep-neural network to a lab. c) The novel feedback-loop training mechanism in FBGAN. At every epoch, several predictions are sampled from the generator and input into the analyzer. The analyzer gives a score to each sequence as demonstrated in b, and the highest scoring sequences are selected. These high scoring sequences are input back into the discriminator as "real" data. $n$ selected sequences from the analyzer replace the $n$ oldest sequences in the "real" training dataset of the discriminator. In this way, gradually the discriminator's set of "real" data is replaced by synthetic data receiving high scores from the analyzer.}
\label{flowchart}
\end{figure}

As shown in Figure \ref{flowchart}, the feedback-loop mechanism consists of two components. The first component is the GAN, which generates novel gene sequences which have not been enriched for any properties. The second component is the analyzer; in our first use case, the analyzer is a differentiable neural network which takes in a gene sequence and predicts the probability that the sequence will code for an antimicrobial peptide (AMP). However, the analyzer can be any black-box which takes in a gene sequence and predicts the desirability of the gene sequence with a certain score. For instance, in our second use case the analyzer is a web server which returns the number of alpha-helical residues a gene will code for. The analyzer could even be a scientist who experimentally validates the produced gene sequences, which would be an example of \textit{active learning}.

The GAN and analyzer are linked by the feedback mechanism after an initial number of pretraining epochs so that the generator is producing valid sequences. Once the feedback mechanism starts, once every epoch a set number of sequences are sampled from the generator and input into the analyzer. The analyzer predicts how favorable each gene sequence is, and the $n$ top favorable sequences are input back into the discriminator as "real" data that the generator must now mimic in order to minimize its loss. The generated sequences replace the oldest $n$ genes that were present in the discriminator's training dataset. The GAN is then trained as usual for one epoch (one pass through this training set). As the feedback process continues, the entire training set of the discriminator is replaced repeatedly by generated sequences that have received high scores from the analyzer.

\subsection{Analyzer for Antimicrobial-Peptide (AMP) Coding Genes}
The analyzer was a classifier whose input was a gene sequence and output was a probability that the gene coded for an AMP.

\subsubsection{Dataset}
The AMP classifier was trained on 2600 experimentally verified antimicrobial peptides from the APD3 database \cite{apd3}, and a negative set of 2600 randomly extracted peptides from UniProt from 10 tO 50 amino acids (filtered for unnatural amino acids). The dataset was loaded using the Modlamp package \cite{modlamp}. As above, the proteins were translated to cDNA by translating each amino acid to a codon (a random codon in the case of redundancy), and by adding a start codon and random stop codon. The AMP training dataset was split into 60\% training, 20\% validation, and 20\% test sequences.

\subsubsection{Classifier Architecture}
Using Pytorch, we built and trained a Recurrent Neural Network (RNN) Classifier to predict whether a gene sequence would produce an antimicrobial peptide (AMP). The architecture of the RNN consisted of two GRU (Gated Recurrent Unit) layers with hidden state h = $128 \times 1$. The second LSTM layer’s output at the final time step was fed to a dense output layer, with the number of neurons equal to the number of output classes minus one. This dense layer had a sigmoid activation function, such that the output corresponded to the probability of the gene sequence being in the positive class. In order to reduce overfitting and improve generalization, we added dropout with $p=0.3$ in both layers.

Using the Adam optimizer with learning rate lr = 0.001, we optimized the binary cross entropy loss of this network. The network was trained using minibatch gradient descent with batch size B = 64, and 60\% of the data was retained for training, 20\% for validation, and 20\% for testing. The model was trained for 30 epochs.

\subsection{Secondary-Structure Black Box Analyzer}
In order to optimize the synthetic genes for secondary structure, a wrapper was written around the PSIPRED predictor of secondary structure \cite{psipred}. The PSIPRED predictor takes in an amino-acid sequence and tags each amino acid in the sequence with known secondary structures, such as alpha-helix or beta-sheet. The wrapper takes in a gene sequence (sampled from the generator), converts it into a protein sequence, and predicts the secondary structure of the amino acids in that protein sequence. The wrapper then output the total number of alpha-helix tagged residues from the sequence. If the gene cannot be converted into a valid protein sequence, the wrapper outputs zero. The analyzer selects all sequences with helix length above some cutoff to move to the discriminator's training set. In this case, the cutoff was arbitrarily set to five residues.

\section{Results and Discussion}
\subsection{WGAN Architecture to Generate Protein-Coding Sequences}
\begin{figure}[]
\centering
    \includegraphics[width=0.8\textwidth]{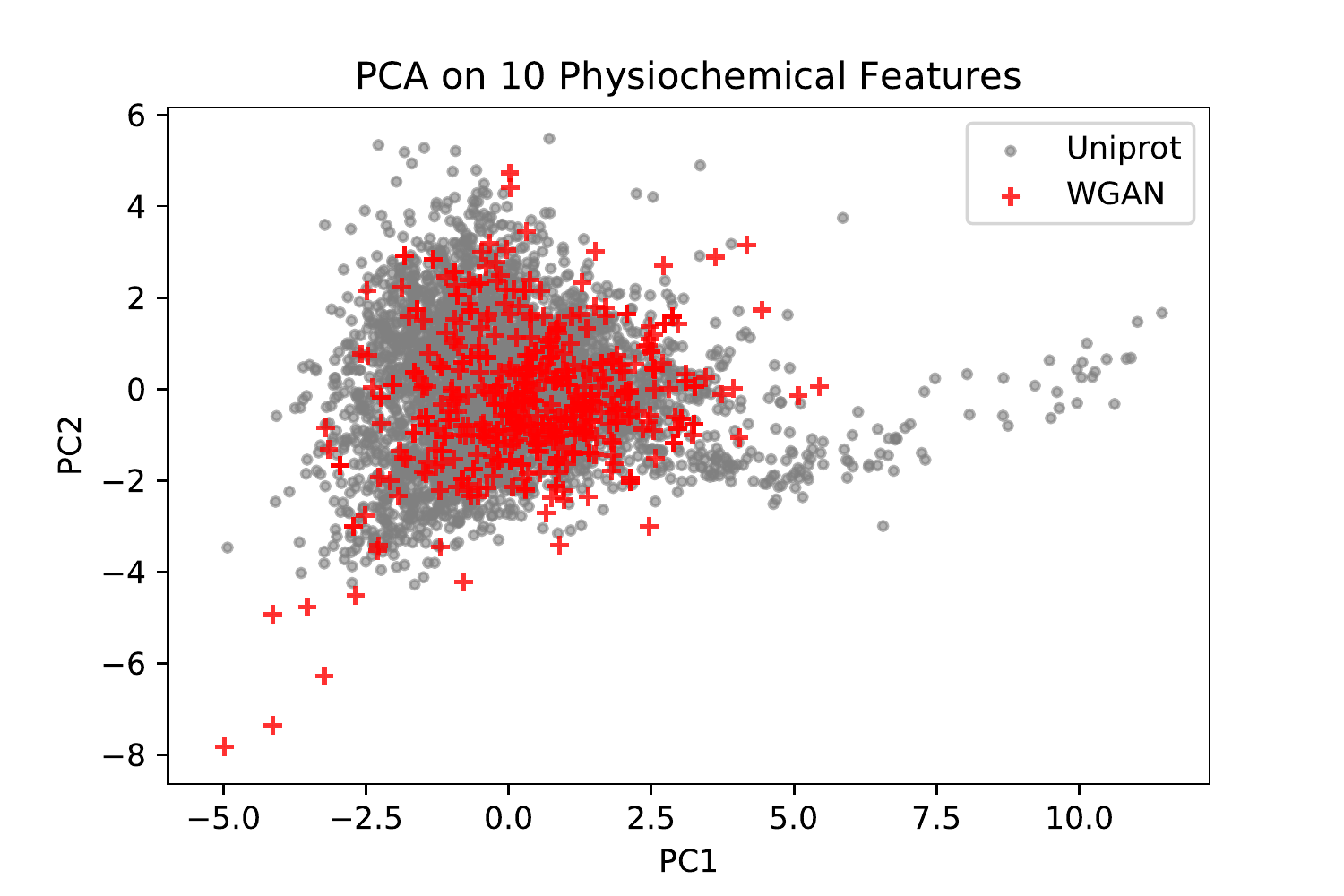}
  \caption{A set of 500 valid genes were sampled from the trained WGAN, and 10 physiochemical features were calculated for the proteins encoded by the synthetic genes. The same 10 features were also calculated for the cDNA sequences from Uniprot proteins. PCA was performed on the features of the natural cDNA sequences, and the synthetic genes were transformed accordingly. The first two principal components (PC1, PC2) are shown here; we see that the synthetic sequences lie in the same chemical space as the natural sequences.}
  \label{pca}
\end{figure}

Synthetic genes up to 156 nucleotides (50 amino acids) in length are produced from the WGAN architecture with gradient penalty; after training, three batches of sequences (192) sequences were sampled from the generator. Three batches were also sampled before one epoch of training.

The correct gene structure was defined as a string starting with the canonical start codon, following with an integer number of codons of length 3, and ending with one of three canonical stop codons. Before training, $3.125\%$ of sequences initially followed the correct gene structure. After training, $77.08\%$ of sampled sequences contained the correct gene structure, demonstrating a large improvement after training.

In order to further examine whether the synthetic genes were similar to natural cDNA sequences extracted from Uniprot, several physiochemical features of the resulting proteins were calculated such as length, molecular weight, charge, charge density, hydrophobicity, etc. These features were calculated for the synthetic genes and the natural cDNA sequences extracted from Uniprot, and a principal component analysis (PCA) was conducted on these physiochemical features. The PCA was fit on the features of the natural cDNA sequences, and the synthetic sequences were transformed accordingly.

\begin{figure}[ht]
 \begin{minipage}{\textwidth}
    a) Uniprot Proteins\\
    \begin{center}
    \includegraphics[width=0.7\textwidth]{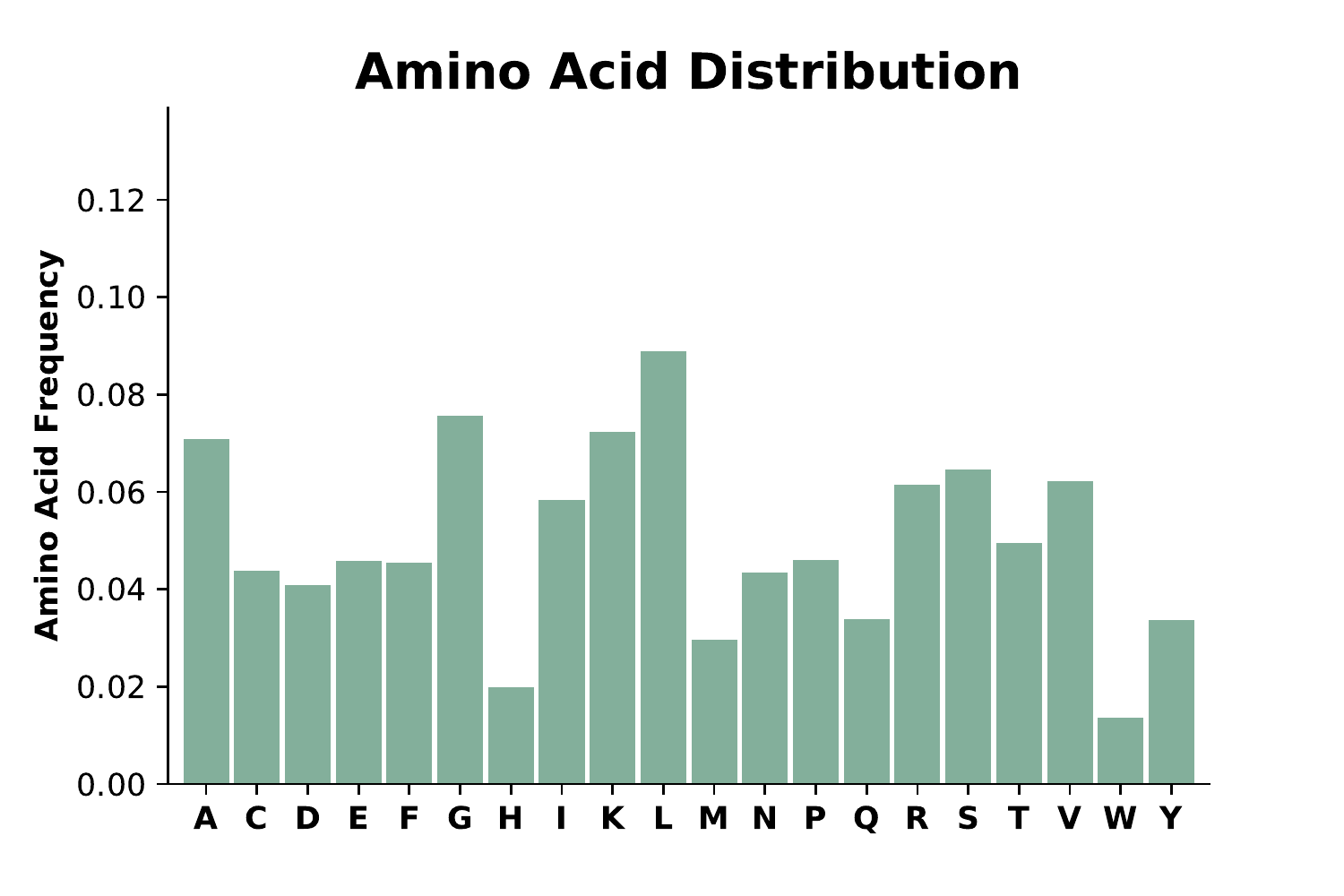}
    \end{center}
  \end{minipage}

  \begin{minipage}{\textwidth}
    b) Synthetic Genes (No Feedback)\\
    \begin{center}
    \includegraphics[width=0.7\textwidth]{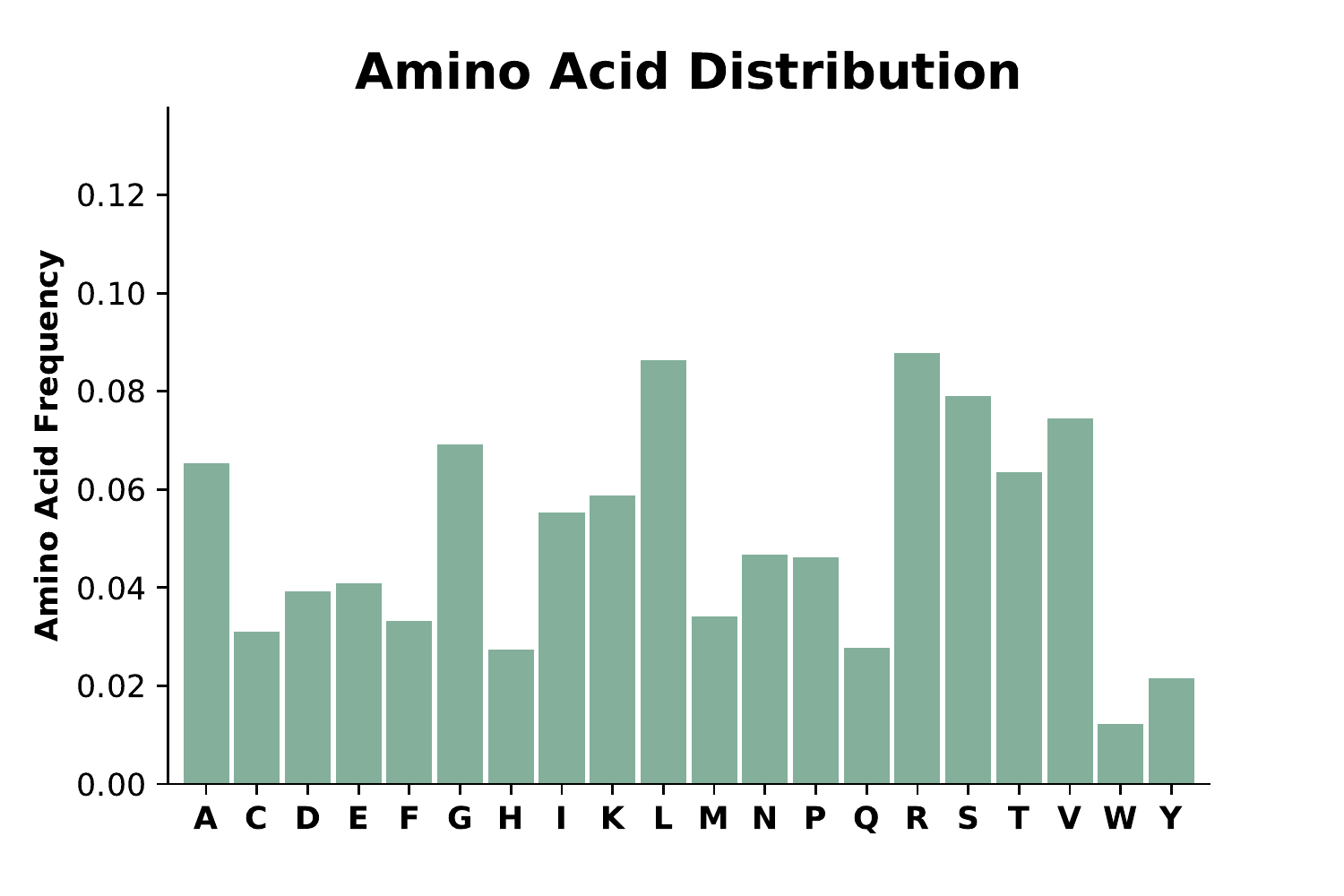}
    \end{center}
  \end{minipage}
  \caption{Amino acid distributions for a) Natural proteins gathered from Uniprot, b) Synthetic genes produced from WGAN. The relative frequencies of the amino acids in the synthetic genes mirror the relative frequencies from the natural Uniprot proteins, providing evidence that the WGAN has learned to reproduce certain properties of natural genes.}
  \label{aa_dist}
\end{figure}

Figure \ref{pca} shows the Uniprot gene sequences and generated genes plotted with respect to these principal components, and we see that both the natural and synthetic sequences lie in the same space. In addition, as shown in Figure \ref{aa_dist}, the relative amino acid frequencies of the synthetic sequences mirror the relative frequencies of the natural cDNA sequences from Uniprot.

\subsection{Deep RNN Analyzer for Antimicrobial Properties}
The AMP analyzer used a Recurrent Neural Network (RNN) to score each gene sequence with its probability of producing an AMP. The architecture of the RNN consisted of two GRU (Gated Recurrent Unit) layers with hidden state h = $128 \times 1$ and dropout $p=0.3$ in both layers.

\begin{figure}[h]
\centering
    \includegraphics[width=0.5\textwidth]{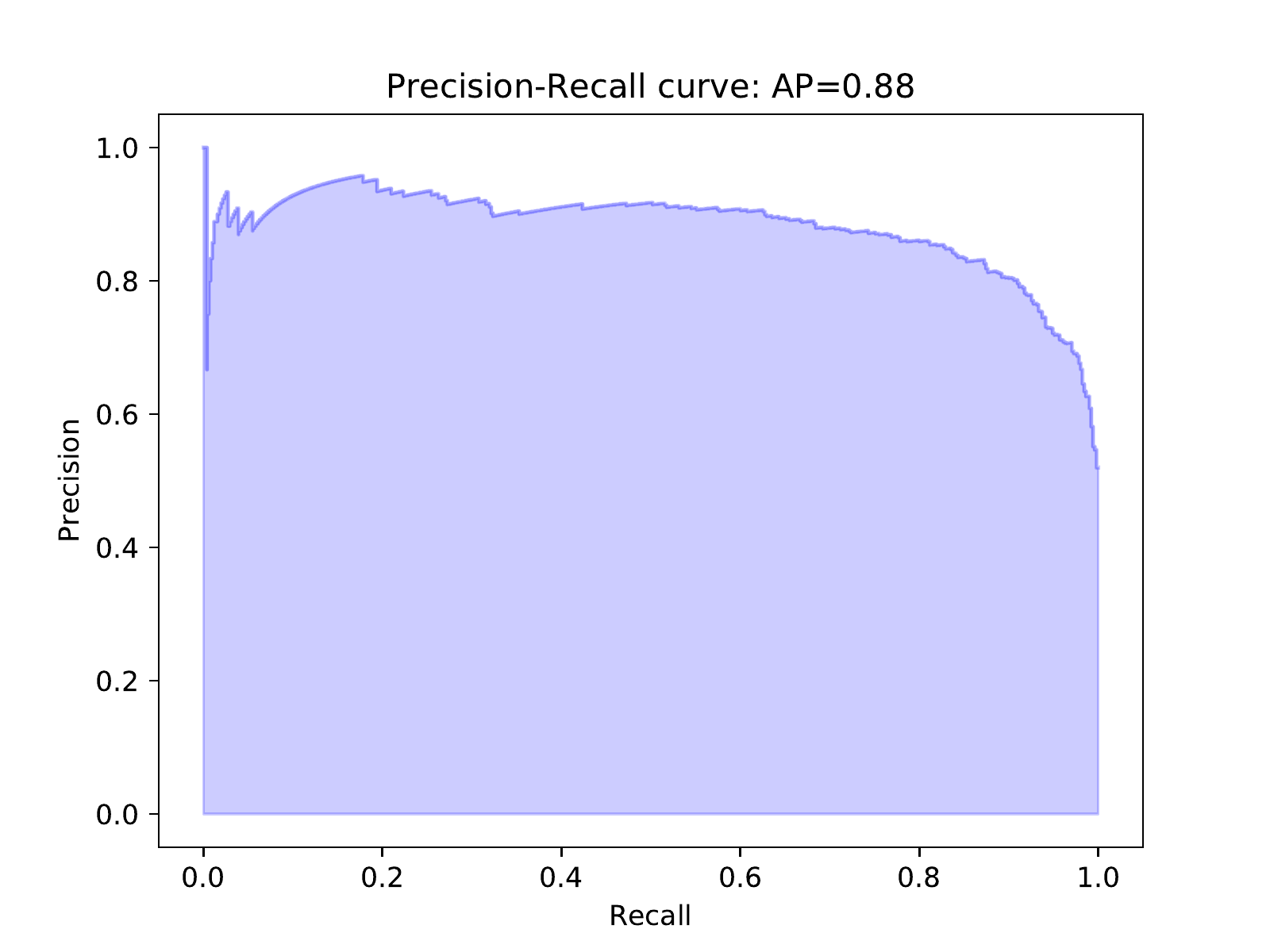}
  \caption{Precision-recall curve for AMP analyzer, which used two GRU layers to preduct whether a given gene would produce an antimicrobial protein. AUPRC was 0.88.}
  \label{AMP_prec_recall}
\end{figure}

In order to quantitatively measure the performance of the classifier, we measured the analyzer's accuracy, AUROC, precision, and recall. The model achieved a training accuracy of $0.9447$ and a validation accuracy of $0.8613$. The test accuracy was $0.842$, and the AUROC on the test set was $0.908$. The precision and recall on the test set were $0.826$ and $0.8608$, respectively, and the area under the precision-recall curve was $0.88$, as shown in Figure \ref{AMP_prec_recall}.

\subsection{Feedback-Loop to Optimize Antimicrobial Properties}
After both the GAN and function analyzer were trained, the two were linked with the described feedback-loop; at each epoch of training, sequences were sampled from the generator and fed into the analyzer. The analyzer then assigned each sequence a probability of being antimicrobial, and the top ranking sequences (here with $P(Antimicrobial) > 0.8$) were fed into the discriminator and labelled as "real" sequences. The $n$ top ranking sequences took the place of the $n$ oldest sequences in the discriminator's data set.

In order to measure the effectiveness of this feedback-loop mechanism in FBGAN, two criteria were examined. The first criteria was whether the analyzer predicted more of the outputs from the generator to be antimicrobial over time (without sacrificing the gene structure); the second criteria was whether the generated genes were similar to known antimicrobial genes, in both their sequences and in the properties of the resulting proteins.

\begin{figure}[]
 \begin{minipage}{\textwidth}
    a)\\
    \begin{center}
    \includegraphics[width=0.7\textwidth]{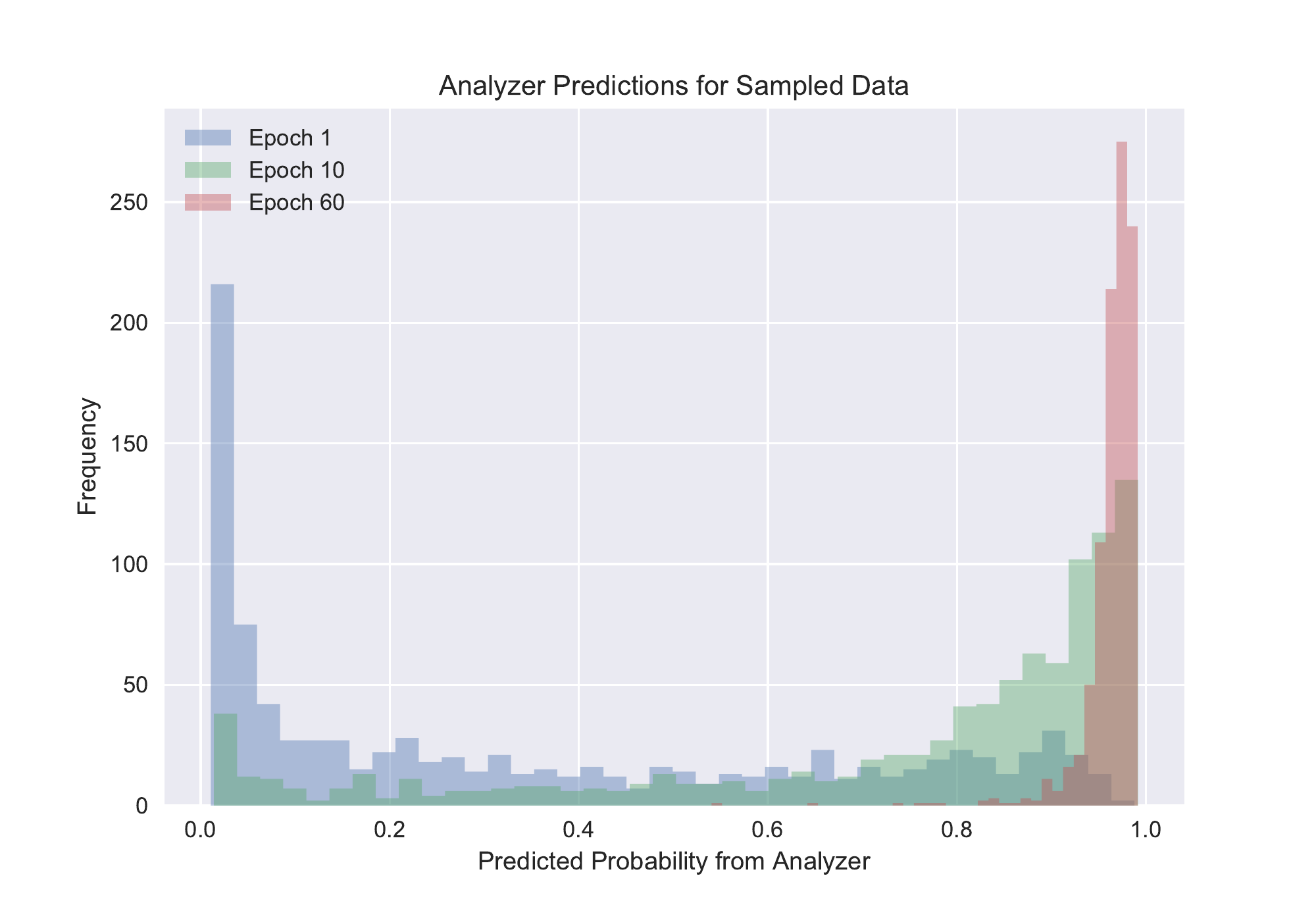}
    \end{center}
  \end{minipage}
  \begin{minipage}{\textwidth}
    b)\\
    \begin{center}
    \includegraphics[width=0.7\textwidth]{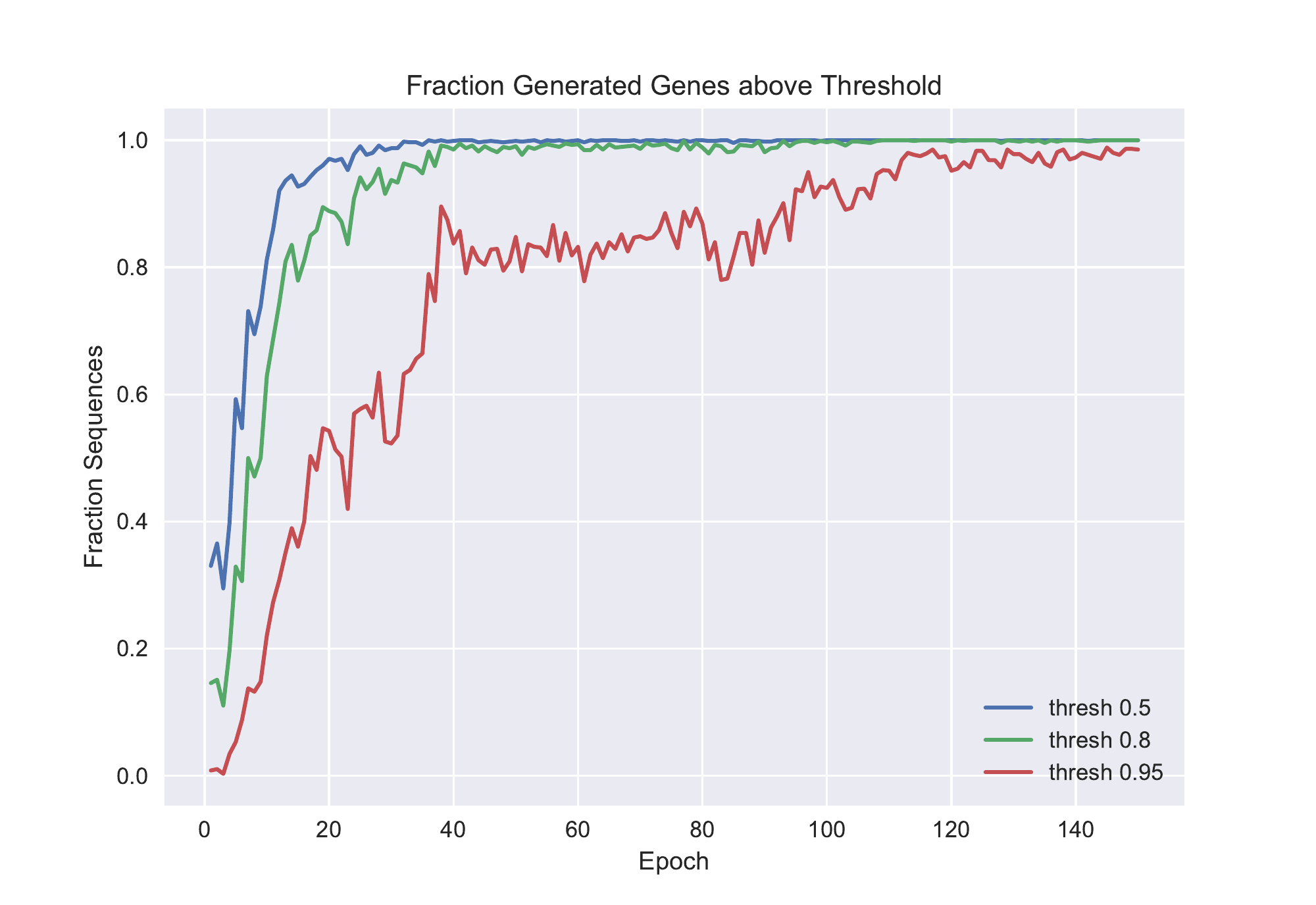}
    \end{center}
  \end{minipage}
  \caption{a) Histograms showing predicted probability that generated genes are antimicrobial, as the closed-loop training progresses. While most sequences are initially assigned $0.1$ probability of being antimicrobial, as training progresses, nearly all sequences are eventually predicted to be antimicrobial with probability $>0.99$. b) Percentage of sequences predicted to be antimicrobial with probability above three thresholds: $[0.5, 0.8, 0.99]$. While $0.8$ was used as the cutoff for feedback, the percentage of sequences above $0.99$ also continues to rise during training with feedback.}
  \label{amp_analyzer_curves}
\end{figure}

In order to answer the first question, we examined the analyzer's predictions on the generator's sequences as the training progressed with feedback. As shown in Figure \ref{amp_analyzer_curves}, after only ten epochs of closed-loop training, the analyzer predicts the majority of sequences as being antimicrobial. After sixty epochs, nearly all the sequences are predicted to be antimicrobial with high probability (greater than $0.99$). Even though the threshold for feedback was at $0.8$, the generator continues to improve even beyond the threshold, suggesting that the closed-loop training is robust to changes in the threshold value. Moreover, 93.3\% of the generated sequences after closed-loop training have the correct gene structure, showing that the reading frame structure was not sacrificed but rather reinforced.

\begin{figure}[h]
\begin{minipage}{0.5\textwidth}
   a)\\
   \includegraphics[width=\textwidth]{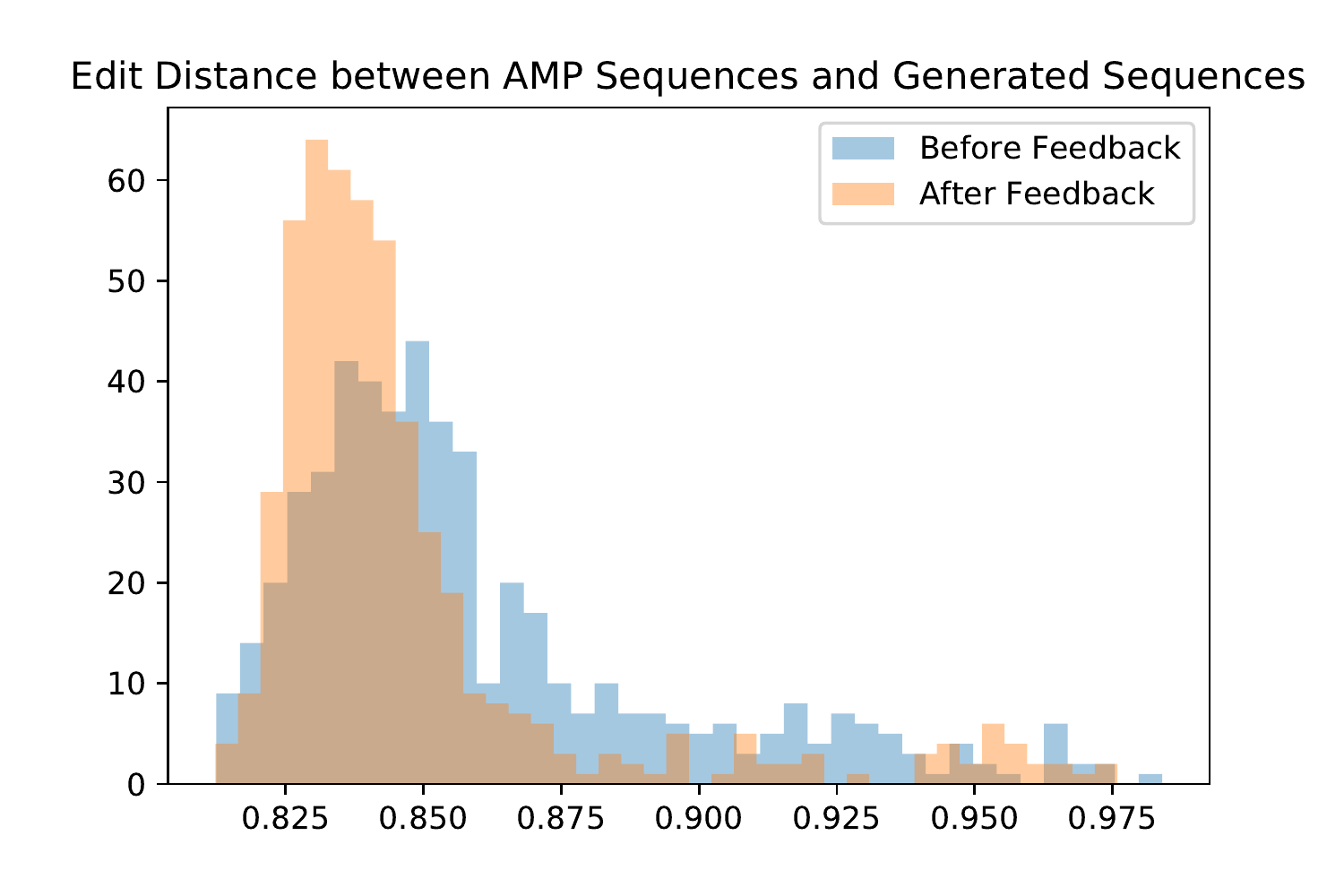}
 \end{minipage}
 \begin{minipage}{0.5\textwidth}
   b)\\
   \includegraphics[width=\textwidth]{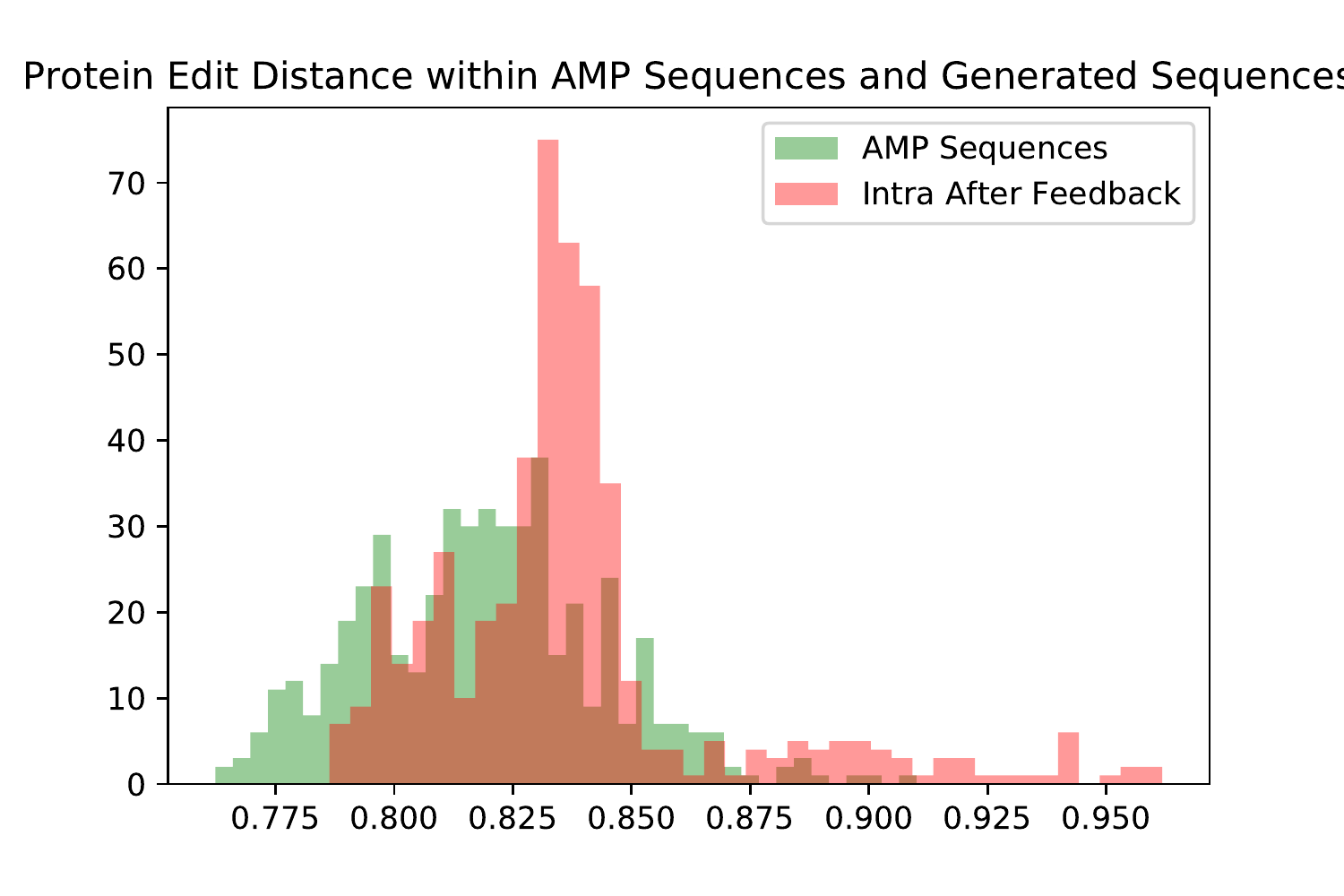}
 \end{minipage}

  \caption{a) Between-group edit distance (Levenstein distance) between known antimicrobial sequences (AMPs) and 1) proteins coded for by synthetic genes produced without feedback, and 2) proteins coded for by synthetic genes produced after feedback. In order to calculate between-group edit distance, the distance between each synthetic protein and each AMP was calculated and the means were then plotted. b) Within-group edit distance for AMPs and for proteins produced after feedback, to evaluate the variability of GAN generated genes after the feedback loop.  Within-group edit distance was computed by selecting 500 sequences from the group and computing the distance between each sequence and every other sequence in the group; the mean of these distances was then taken and plotted.}
  \label{edit_distance}
\end{figure}

Next the generated sequences were examined for similarity with the experimental antimicrobial genes, according to both sequence and physiochemical properties of the proteins coded for. Figure \ref{edit_distance}a shows a histogram of the mean edit distance between the known AMPs and proteins from synthetic genes before feedback, and the distance between the AMPs and proteins from synthetic genes produced after feedback. Figure \ref{edit_distance}b shows the intrinsic edit distance within the AMP proteins, and within the proteins coded for by the synthetic gene sequences after feedback. All edit distances were normalized by the length of the sequences, in order to not penalize longer sequences unfairly.

The distribution of edit distances shifts after feedback to have a larger proportion of sequences with a lower edit distance from the AMP sequences. In addition, the sequences after feedback have a higher edit distance within themselves than the antimicrobial sequences do with each other; this demonstrates that the model has not overfit to replicate a single data point.

Next the physiochemical properties of the resulting proteins were measured, and are shown in Table 1. As can be seen in the table, the proteins encoded by the closed-loop sequences shift to be closer to the positive antimicrobial peptides in five out of ten physiochemical properties such as Length, Hydrophobicity, and Aromaticity, and remains as similar as the sequences without feedback for properties such as Charge and Aliphatic index. This is true even though the analyzer operated directly on the gene sequence rather than these physiochemical properties, and so the feedback mechanism did not directly optimize the physiochemical properties that show a shift.

\begin{table}[h]
\begin{tabular}{l|l|l|l}
& Positive AMP         & Before Feedback & After Feedback \\ \hline
\textbf{Length}              & 32.37 $\pm$ 17.983               & 21.419 $\pm$ 13.190                & 36.992$\pm$ 16.978      \\
\textbf{Molar Weight}       & 3514.0068 $\pm$ 1980.59          & 2419.032 $\pm$ 1479.013            & 4023.584 $\pm$ 1848.048 \\
Charge               & 3.8575 $\pm$ 2.979               & 2.356 $\pm$ 2.447                  & 2.708 $\pm$ 2.249       \\
Charge Density       & 0.00123 $\pm$ 0.00084            & 0.00127 $\pm$ 0.00138              & 0.00091 $\pm$ 0.00096   \\
pI                   & 10.2697 $\pm$ 2.046              & 10.143 $\pm$ 2.444                 & 9.474 $\pm$ 1.844       \\
Instability Index    & 27.174 $\pm$ 26.717              & 37.791 $\pm$ 35.697                & 53.145 $\pm$ 29.495     \\
\textbf{Aromaticity  }        & 0.0822  $\pm$ 0.0602             & 0.0642 $\pm$ 0.0695                & 0.0775 $\pm$ 0.066      \\
Aliphatic Index      & 91.859 $\pm$ 47.236              & 84.397 $\pm$ 45.681                & 84.889 $\pm$ 34.837     \\
\textbf{Boman Index }         & 0.770 $\pm$ 1.500                & 1.801 $\pm$ 1.721                  & 0.888 $\pm$ 1.155       \\
\textbf{Hydrophobicity Ratio} & 0.435 $\pm$ 0.128                & 0.390 $\pm$ 0.144                  & 0.441 $\pm$ 0.109
\end{tabular}
\begin{center}
\includegraphics[width=\textwidth]{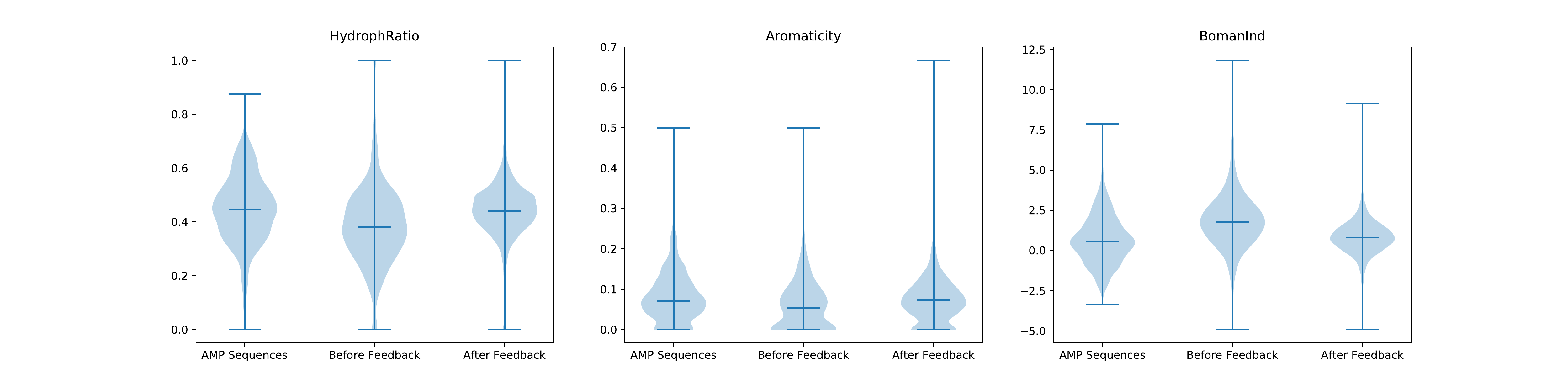}
\label{properties_amp}
\end{center}

\caption{Mean $\pm$ standard deviation of physiochemical features, before and after feedback. Bolded properties are those for which the mean after feedback is closer to the mean of the positive sequences than the mean before feedback. Violin plots of some example properties demonstrating the shift after feedback are shown below.}
\end{table}

\subsection{Optimizing Secondary Structure with Black-Box PSIPRED Analyzer}
The generator was then optimized to produce synthetic genes with a particular secondary structure in their products, in this case alpha-helical peptides. Besides being extremely important for protein function, secondary structure is attractive to optimize for since it arises in short peptides of length less than 50, such as the sequences being generated here.

The analyzer used to optimize for helical peptides was a black-box secondary structure predictor from the PSIPRED server, which tags protein sequences with predicted secondary structure \cite{psipred} at each amino acid. All gene sequences with more than 5 alpha-helical residues were input back into the discriminator as real data.

\begin{figure}[h]
\includegraphics[width=0.5\textwidth]{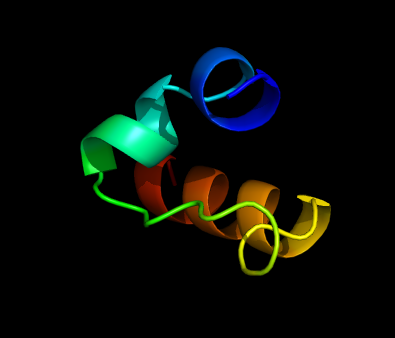}
\includegraphics[width=0.5\textwidth]{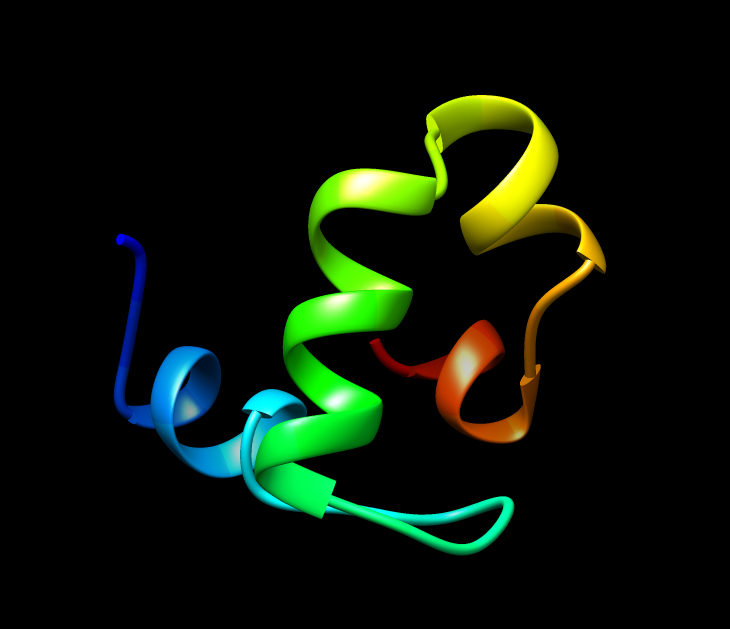}
\caption{Example peptides from the synthetic genes output by our WGAN model with feedback from the PSIPRED analyzer. Both proteins show a clear helix structure. The peptide on the left was predicted to have 10 residues arranged in helices, while the peptide on the right was predicted to have 22 resides in helices; accordingly, the peptide on the right appears to have more residues arranged in helices. }
\label{peptide_models}
\end{figure}

\begin{figure}[h]
 \begin{minipage}{\textwidth}
    a)\\
    \begin{center}
    \includegraphics[width=0.8\textwidth]{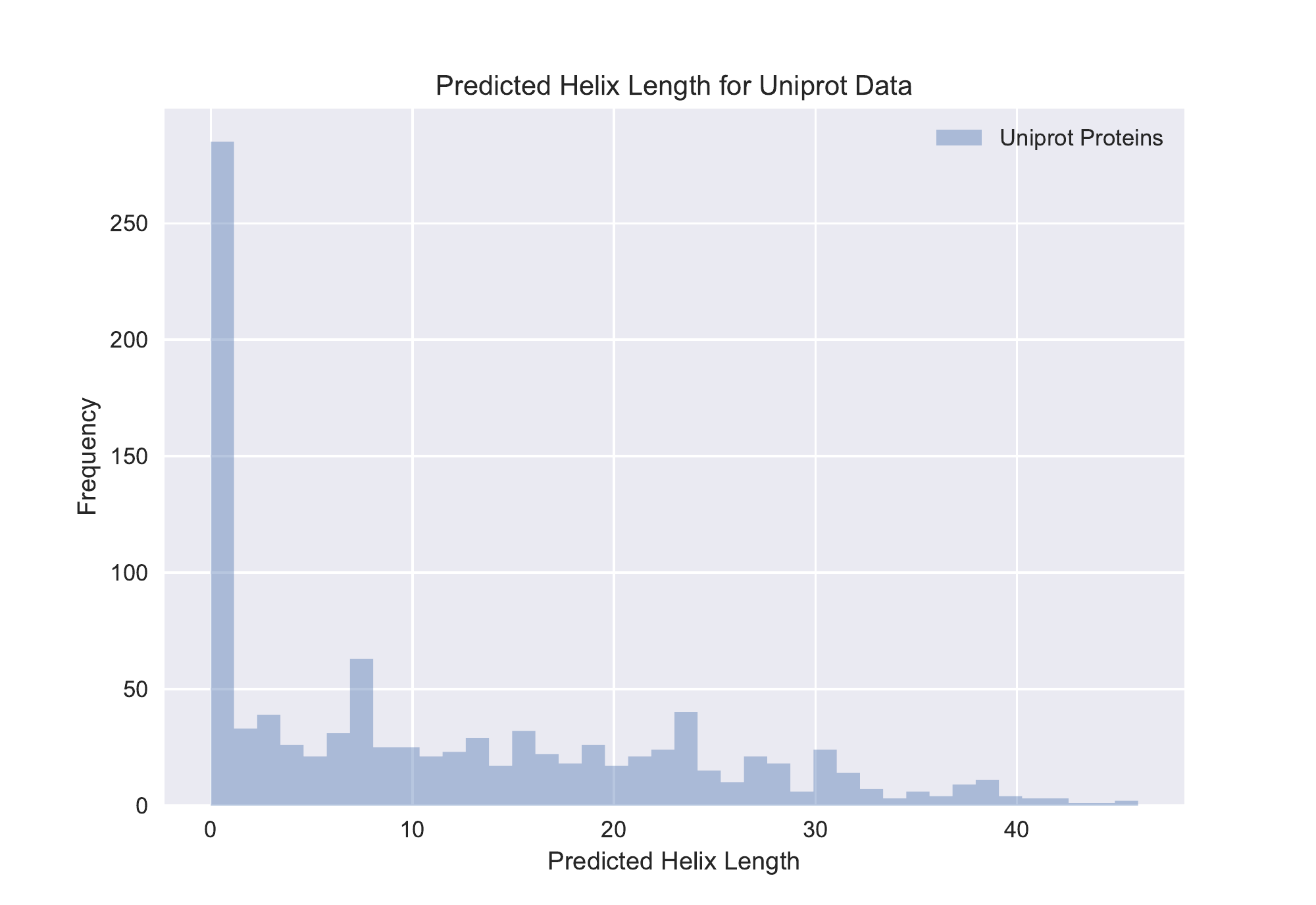}
    \end{center}
  \end{minipage}

  \begin{minipage}{\textwidth}
    b)\\
    \begin{center}
    \includegraphics[width=0.8\textwidth]{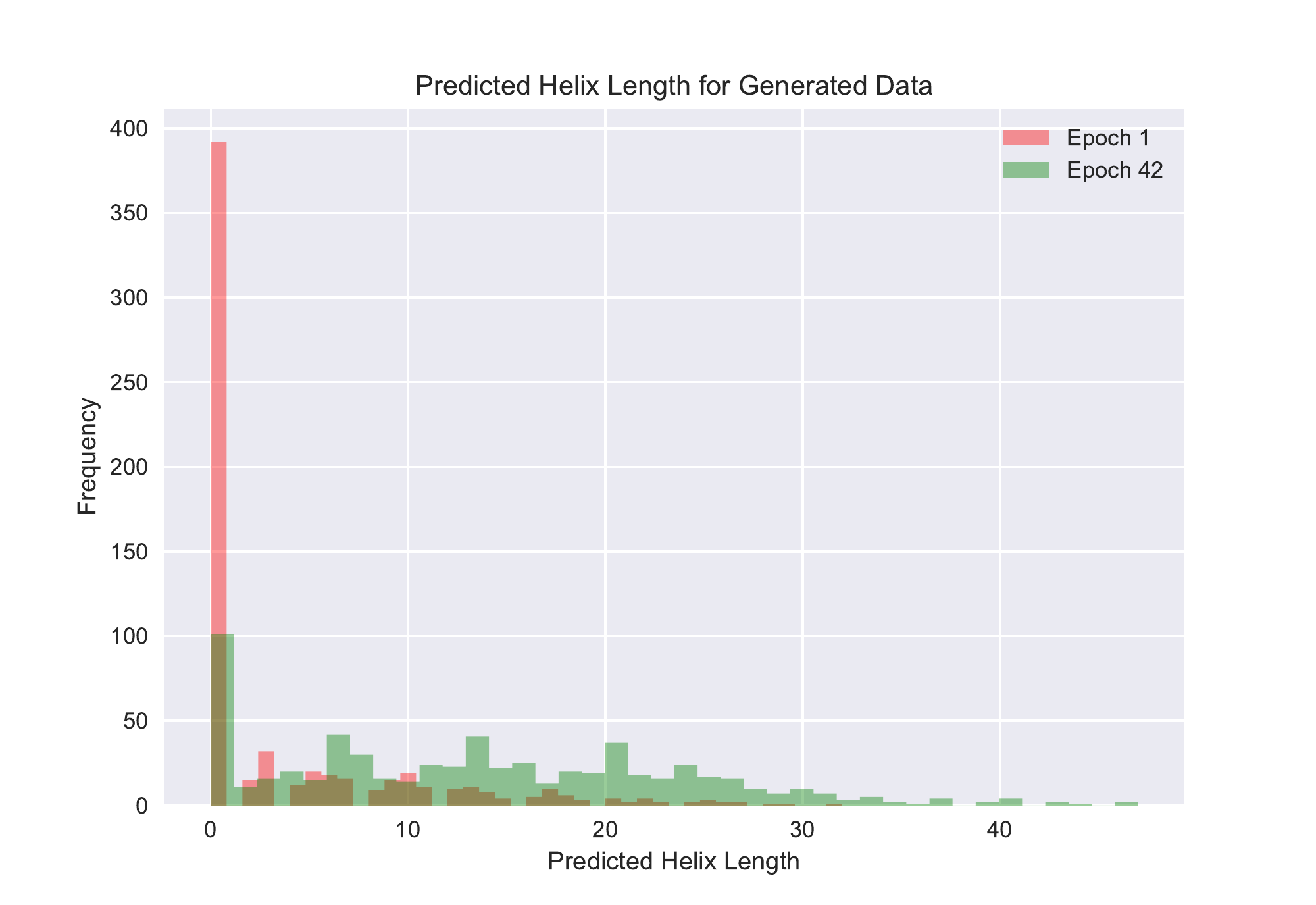}
    \end{center}
  \end{minipage}
  \caption{a) Distribution showing alpha-helix lengths for natural proteins under 50 amino acids scraped from Uniprot. b) Distribution of alpha-helix lengths from synthetic gene sequences after 1 and 40 epochs of training. The predicted helix length from the generated sequences quickly shifts to be higher than the helix length of the natural proteins.}
  \label{helix_len}
\end{figure}

\begin{figure}
\centering
\includegraphics[width=0.7\textwidth]{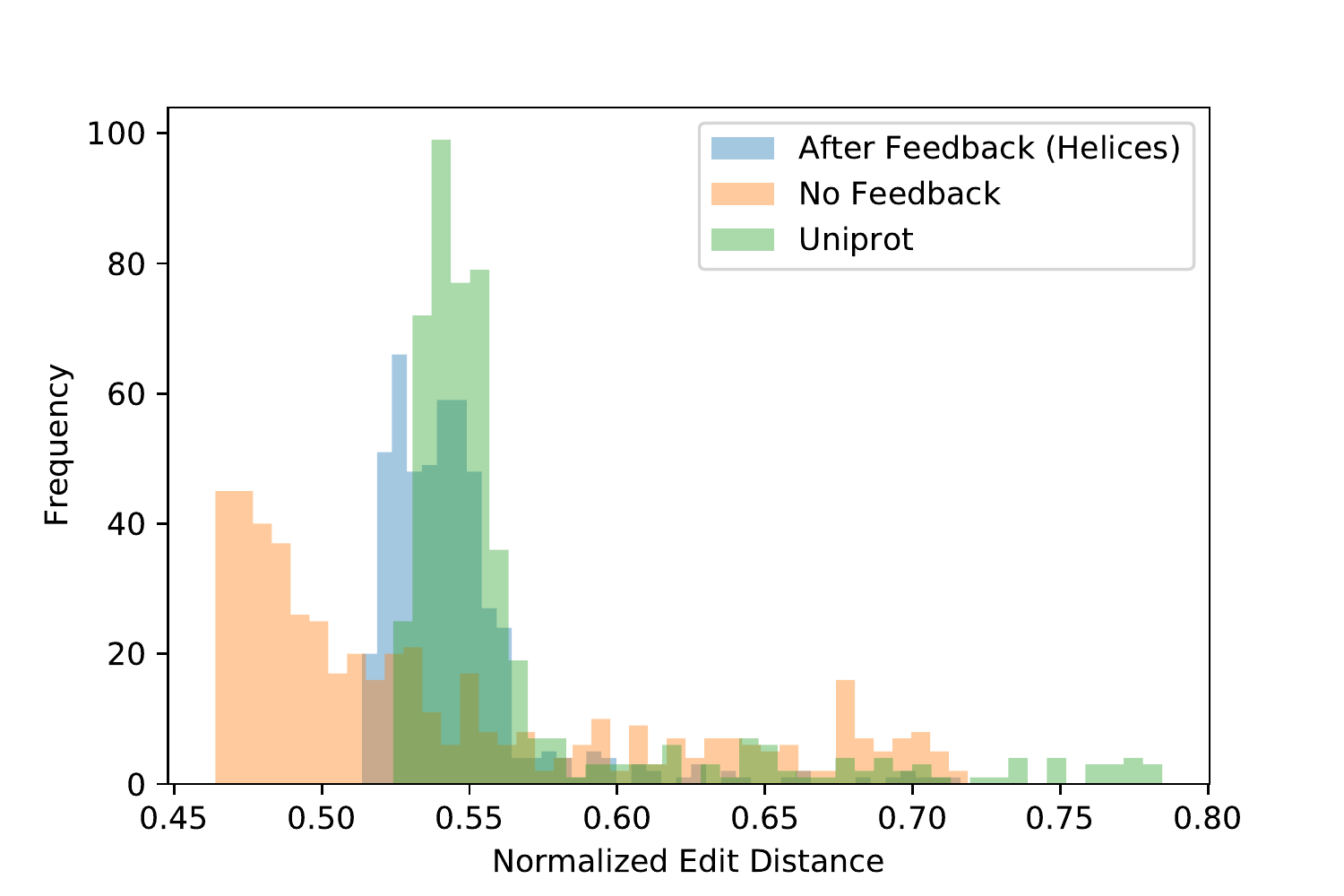}
\caption{Within-group edit distance for natural cDNA sequences from Uniprot, for synthetic genes before feedback, and after feedback from the PSIPRED secondary-structure analyzer. Within-group edit distance was computed by selecting 500 sequences from the group and computing the distance between each sequence and every other sequence within the group; the mean of these distances was then taken and plotted here.}
\label{supp_edit_dist}
\end{figure}

After 43 epochs of feedback, the helix length in the generated sequences was significantly higher than the helix length without feedback and the helix length of the original Uniprot proteins, as illustrated by Figure \ref{helix_len}. Folded examples of peptides we generated are shown in Figure \ref{peptide_models}; these 3D peptide structures were produced by \textit{ab initio} folding from our generated gene sequences, using knowledge-based force field template-free folding from the QUARK server \cite{quark}. The edit distance within the DNA sequences generated after PSIPRED feedback was in the same range as the edit distance within the Uniprot natural cDNA sequences, and higher than the edit distance within the synthetic sequences generated before feedback \ref{supp_edit_dist}.

\section{Conclusion and Future Work}

In this work, we have successfully developed a GAN model, FBGAN, to produce novel protein-coding sequences for peptides under 50 amino acids in length, and demonstrated a novel feedback-loop mechanism to optimize those sequences for desired properties. We use a function analyzer to evaluate sampled sequences from the generator at every epoch, and input the best scoring sequences back into the discriminator as "real" data points. In this way, the outputs from the generator slowly shift over time to outputs that are highly predicted to be positive by the function analyzer. This feedback-loop mechanism, to our knowledge, has not been proposed before for use in GANs; we have shown that this training mechanism is robust to the type of analyzer used, as the analyzer need not be a deep neural network in order for the feedback mechanism to be successful.

We have demonstrated the usefulness of the feedback-loop mechanism in two use cases: 1) optimizing for genes that code for antimicrobial peptides (AMPs), and 2) optimizing for genes that code for alpha-helical peptides. For the first use case, we built our own deep RNN analyzer; for the second, we employed the existing PSIPRED analyzer in a black-box manner. In both cases, we were able to significantly shift the generator to produce genes likely to have the desired properties.

It is useful to have the ability to optimize synthetic data for desired properties without a differentiable analyzer for two reasons: first of all, this allows the analyzer to be any model that takes in a datapoint and assigns it a score; the analyzer may now even be a machine carrying out experiments in a lab. The second reason is that many existing models in bioinformatics are based on non-differentiable operations, such as BLAST searches or homology detection algorithms. This feedback loop mechanism thus allows previous staples of synthetic biology research to integrate smoothly with the enormous capabilities of GANs. The feedback-loop technique is also desirable precisely because it is robust, simple, and easy to implement.

While we were able to extend the GAN architecture to produce genes up to 156 base pairs in length while maintaining the correct start codon/stop codon structure, it was noticeably more difficult to maintain the structure at 156 base pairs than at 30 base pairs or 50. In order to allow the generator to learn patterns in the sequence over longer lengths, we might investigate using a recurrent architecture or even dilated convolutions in the generator, which have been shown to be effective in identifying long-term genomic dependencies \cite{Gupta_dilatedCNN}. It is still challenging to use GAN architectures to produce long, complex sequences, which currently limits the usefulness of GANs in designing whole proteins, which can be thousands of amino acids long.

Here, in order to make the training process for the GAN easier, we focus on producing gene sequences which have a clear start/stop codon structure and only four nucleotides in the vocabulary. However, in the future, we might focus on producing protein sequences directly (with a vocabulary of 26 amino acids).

While we have shown that the proteins from the synthetic genes have shifted after training to be more physiochemically similar to known Antimicrobial peptides, we would like to conduct additional experimental validation on the generated peptides. The same holds true for the predicted alpha-helical peptides.

In future work, we would also like to apply and further validate the currently proposed method on additional application areas in genomics and personalized medicine, such as noncoding DNA and RNA. In addition, FBGAN's proposed feedback-loop mechanism for training GANs is not limited to sequences or to synthetic biology applications; thus, our future work also includes applying this methodology to image generation use-cases of GANs.

\bibliography{sample}

\beginsupplement

\subsection{Appendix}

\begin{figure}[h]
\centering
\includegraphics[width=0.7\textwidth]{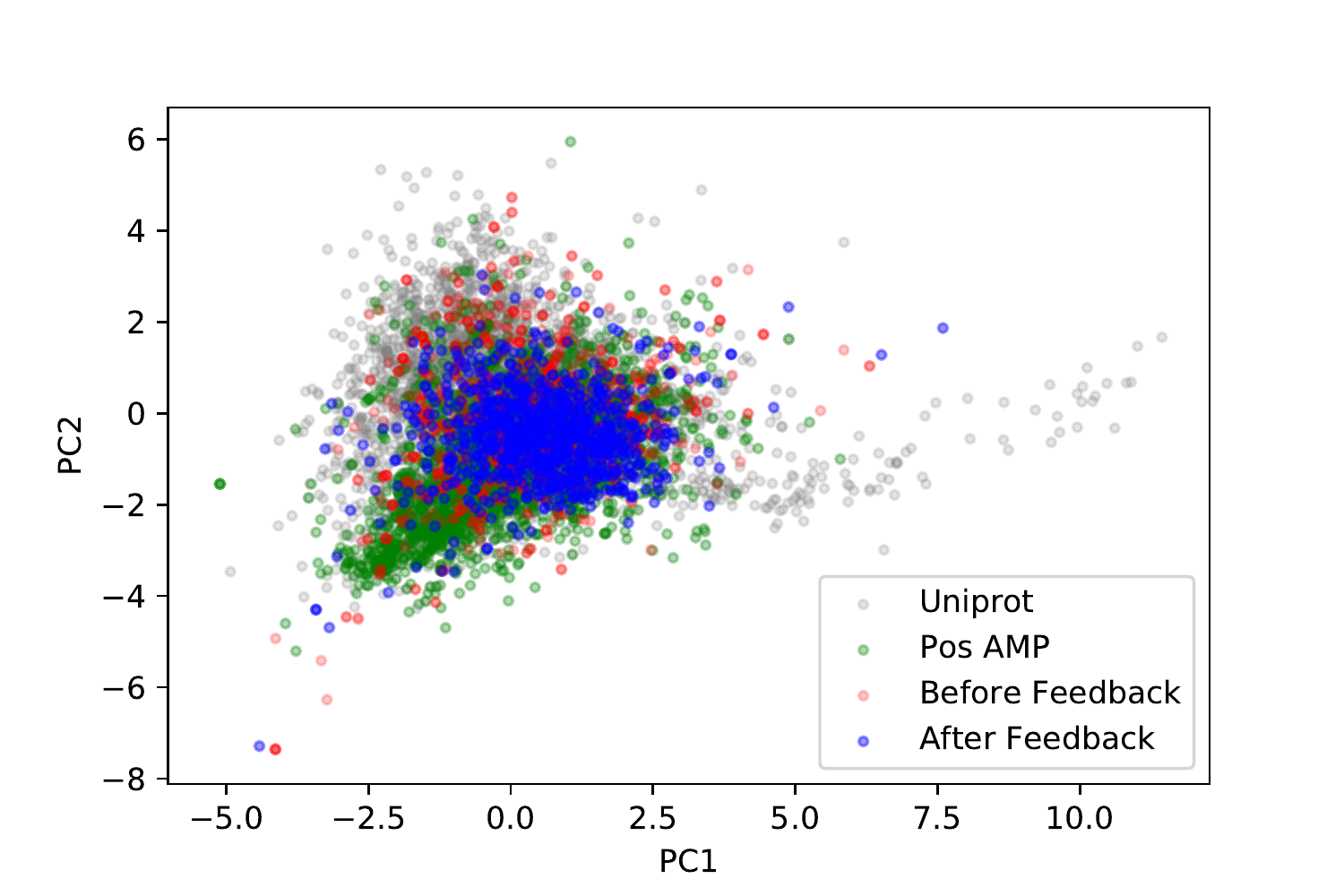}
  \caption{A PCA was conducted based on the 10 physiochemical properties calculated for the natural cDNA sequences, and all other groups of sequences were transformed accordingly. The first two principal components are shown above. The positive AMP sequences lie in a very similar subspace to the Uniprot sequences with respect to these ten physiochemical properties, as do the generated sequences without analyzer feedback. After feedback, the synthetic sequences shrink to the area of greatest density of positive AMP sequences; this makes sense as this is the area of sequences which the analyzer would be expected to predict as antimicrobial with the greatest confidence.}
  \label{feedback_pca}
\end{figure}

\end{document}